\documentclass{PoS}

\title{Limits on the effective quark radius
        from inclusive ep scattering \&
        contact interactions at HERA}

\ShortTitle{Limits on quark radius and contact interactions at HERA}

\author{\speaker{Aleksander Filip \.Zarnecki} ~~on behalf of the ZEUS Collaboration\\
        Faculty of Physics, University of Warsaw\\
        E-mail: \email{Filip.Zarnecki@fuw.edu.pl}}

\abstract{
The high--precision HERA data allow searches for up to TeV scales ``Beyond
the Standard Model'' contributions to electron--quark scattering.
Combined H1 and ZEUS measurements of inclusive deep inelastic cross
sections in neutral and charged current ep scattering are considered,
corresponding to a luminosity of around 1 fb$^{-1}$.
A new approach to the beyond the Standard Model analysis of the inclusive
$ep$ data is presented; simultaneous fits of parton distribution functions
and contributions of ``new physics'' processes are performed.
Considered are possible deviations from the Standard Model due to
a finite radius of quarks, described within the quark form-factor model,
and due to new electron-quark interactions in the framework of
$eeqq$ contact interactions (CI). The resulting 95\% C.L. upper limit
on the effective quark radius is $0.43\cdot 10^{-16}$~cm. The limits
on the CI mass scale extend up to 10 TeV depending on the CI scenario.
}

\FullConference{38th International Conference on High Energy Physics\\
		3-10 August 2016\\
		Chicago, USA}

\begin{document}

%----------------------------------------------------------------------------

\section{Introduction}

Precision measurements of deep inelastic $e^\pm p$ scattering (DIS)
cross sections at high values of negative four-momentum-transfer
squared,  $Q^2$, allow searches for contributions beyond the Standard
Model (BSM), even far beyond the centre-of-mass energy of the $e^\pm
p$ interactions. 
For many ``new physics'' scenarios, cross sections can be affected
by new kinds of interactions in which virtual BSM particles are
exchanged. 
As the HERA kinematic range is assumed to be far below the scale of
any new physics, all such interactions can be approximated as
contact interactions (CI).

The H1 and ZEUS collaborations measured inclusive 
$e^{\pm}p$ scattering cross sections at HERA from
1994 to 2000 (HERA I) and from 2002 to 2007 (HERA II), 
collecting together a total integrated luminosity of about 1\,fb$^{-1}$.
All inclusive data were combined \cite{h1zeus_inc} to create
one consistent set of neutral current (NC) and charged current (CC)
cross-section measurements for $e^{\pm}p$ scattering with unpolarised beams.
The inclusive cross sections were used as input to a QCD analysis
within the DGLAP formalism, resulting in a PDF set
denoted as \mbox{HERAPDF2.0}.

For the results presented in this contribution a new approach to the
beyond the Standard Model analysis was used, 
based on the simultaneous fits of parton distribution functions and
contributions of the ``new physics'' processes.
This is the only method to properly take into account the possibility that
the PDF set may already have been biased by partially 
or totally absorbing previously unrecognised BSM contributions.

% ----------------------------------------------------------------------------
%       Quark form factor
% ----------------------------------------------------------------------------

\section{New physics scenarios}

One of the possible parameterisations of deviations from SM predictions
in $ep$ scattering is achieved by assigning an effective finite radius to 
electrons and/or quarks.
The expected modification of the SM cross section can be described using
a semi-classical form-factor approach~\cite{Kopp:1994qv}.
For small deviations, the SM predictions for the 
cross sections are modified, approximately, to:
\begin{eqnarray}
\frac{d\sigma}{dQ^{2}} & = & 
\frac{d\sigma^{\rm SM}}{dQ^{2}} \;
\left( 1 - \frac{R_{e}^{2}}{6} \, Q^{2} \right)^{2} \;
\left( 1 - \frac{R_{q}^{2}}{6} \, Q^{2} \right)^{2} \; ,   \label{eq:rq}
\end{eqnarray}
where $R_{e}^{2}$ and $R_{q}^{2}$ are the mean-square radii of the 
electron and the quark, respectively,
related to new BSM energy scales.
In the present analysis, only the possible finite spatial distribution 
of the quark was considered and the electron was assumed to be point-like 
($R_{e}^{2} \equiv 0$). 
Both positive and negative values of $R_{q}^{2}$ were considered.

%
% ----------------------------------------------------------------------------
%  CI models 
% ----------------------------------------------------------------------------
%

Four-fermion contact interactions (CI) represent an effective theory 
which describes low-energy effects  due to physics at much higher 
energy scales.
CI model the effects of heavy leptoquarks,
additional heavy weak bosons and electron or quark compositeness.
The CI approach is not renormalizable 
and is only valid in the low-energy limit.
The vector contact-interaction currents considered here
are represented by additional terms in the Standard Model Lagrangian:
\begin{eqnarray}
{\cal L}_{CI} & = & 
   \sum_{^{i,j=L,R}_{q=u,d,s,c,b}} 
   \eta^{eq}_{ij} (\bar{e}_{i} \gamma^{\mu} e_{i} )
                 (\bar{q}_{j} \gamma_{\mu} q_{j}) \; ,
	\label{eq-cilagr}
\end{eqnarray}
where the sum runs over electron and quark helicities
and quark flavors.
The couplings  $\eta^{eq}_{ij}$ describe 
the helicity and flavor structure of the contact interactions.
It was assumed that the same coupling structure applies to all quarks.
The one-parameter scenarios considered in the presented study
are defined by sets of four coefficients, $\epsilon_{ij}$,
as shown in Tab.~\ref{tab-ci},
and the coupling strength, $\eta$, or compositeness scale, $\Lambda$:
\begin{eqnarray}
  \eta^{eq}_{ij} & = &
  \epsilon_{ij} \;  \eta \; = \;
  \epsilon_{ij} \;  \frac{4 \pi}{\Lambda^{2}} \; .
\end{eqnarray}
Models that differ in the overall sign of the
coefficients $\epsilon^{eq}_{ij}$ are distinct because of the
interference with the SM.
In two scenarios listed in the upper part of Tab.~\ref{tab-ci} parity
is violated, while it is conserved in the remaining scenarios.

% ----------------------------------------------------------------------------
%       QCD+CI analysis
% ----------------------------------------------------------------------------

\section{QCD+BSM fit procedure}
\label{sec-fit}

The QCD analysis method used for the HERAPDF2.0 determination~\cite{h1zeus_inc}
was extended to take into account the possible BSM contributions
to the expected cross section values, as described in the published ZEUS $R_q$
analysis \cite{zeus_rq_paper}.
The PDFs of the proton are described
at a starting scale of $1.9$ GeV$^2$ in terms of $N_{par} = 14$ parameters.  
These parameters, denoted $p_k$ in the following (or $\vec{p}$
for the set of parameters), together with the possible contribution of
BSM phenomena (described by the effective quark-radius squared,
$R_{q}^{2}$, or the CI coupling $\eta$) are fit to the data using a
$\chi^2$ method. For CI, the $\chi^2$ formula is given by:
\begin{equation}
 \chi^2 \left(\vec{p},\vec{s},\eta \right) = %\\
%~~~=
 \sum_i
 \frac{\left[m^i
+ \sum_j \gamma^i_j m^i s_j  - {\mu_{0}^i} \right]^2}
{\left( \textstyle \delta^2_{i,{\rm stat}} +
\delta^2_{i,{\rm uncor}} \right) \,  (\mu_{0}^i)^2}
 + \sum_j s^2_j ~~.
\label{eq:qcdfit}
\end{equation} 
Here, $\mu_{0}^{i}$ and $m^i$ are the respective measured cross-section values
and pQCD+BSM cross-section predictions at the point $i$.
The quantities $\gamma^{i}_j $, $\delta_{i,{\rm stat}} $ and 
$\delta_{i,{\rm uncor}}$ are the relative correlated 
systematic, relative statistical and relative uncorrelated 
systematic uncertainties of the input data. 
The components, $s_j$, of the vector $\vec{s}$ represent the
correlated systematic shifts of the cross sections (given in units of
$\gamma^{i}_j $), which are fit to the data together with the PDF parameter set
$\vec{p}$ and the CI coupling $\eta$. 
The summations extend over all data points $i$
and all correlated systematic uncertainties $j$.
The $\chi^2$ definition used for this analysis results 
in a $\chi^2/$d.o.f. of 1.2 for the SM (N$_{\rm{d.o.f.}}=1131$),
like already observed for HERAPDF2.0 NLO fit~\cite{h1zeus_inc}.
%

% ----------------------------------------------------------------------------
%       Limit setting procedure
% ----------------------------------------------------------------------------

\section{Limit-setting procedure}
\label{sec-limit}

The limits on the BSM contributions are derived
in a frequentist approach using the technique of
replicas \cite{zeus_rq_paper,ci_note}.
Replicas are sets of cross-section values that are
generated by varying all cross sections randomly according to their
known uncertainties.
For the analysis presented here, multiple replica sets were used, each 
covering cross-section values on all points of the $x,Q^2$ grid
used in the QCD fit.
For an assumed true value of the quark-radius squared,
$R_q^{\rm 2\;True}$,  or of the CI coupling, $\eta^{\rm True}$,
replica data sets were created by taking
the reduced cross sections calculated from the nominal PDF fit
(with no BSM contribution) and scaling them
with the cross section ratio given by the quark form factor
or the CI cross section formula.
This resulted in a set of cross-section values $m_{0}^{i}$,
which were then varied randomly within statistical
and systematic uncertainties taken from the data, taking correlations
into account.
All uncertainties were assumed to follow Gaussian
distributions.

Details of the limit setting procedure are described in this paragraph
for the case of the quark form-factor model.
The value of the quark-radius squared determined by the fit to the
data themselves, $R_q^{\rm 2\;Data}$, was taken as a test statistic,
to which values from fits to replicas, $R_q^{\rm 2\;Fit}$, were compared.
To set limits, MC replica cross-section sets for each value 
of $R_{q}^{\rm 2\;True}$ were used for a QCD fit with the PDF parameters and 
the quark radius as free parameters.
The probability to obtain a $R_{q}^{\rm 2 \; Fit}$ value
smaller than that obtained for the actual data, 
Prob($R_{q}^{\rm 2 \; Fit} < R_q^{\rm 2\; Data}$), was then studied as
a function of $R_q^{\rm 2\;True}$, as illustrated in
Fig.~\ref{fig-rq_prob_central}. 
Positive (negative) $R_{q}^{\rm 2\;True}$  values that,
in more than 95\% of the replicas, result in the fitted radius squared value,
$R_{q}^{\rm 2\;Fit}$, greater than (less than) that obtained for the data,
$R_q^{\rm 2\; Data}$, were excluded at the $95\%$~C.L.

\section{Results}
\label{sec-results}

\begin{figure}[p]
\begin{center}
\includegraphics[width=0.56\textwidth]{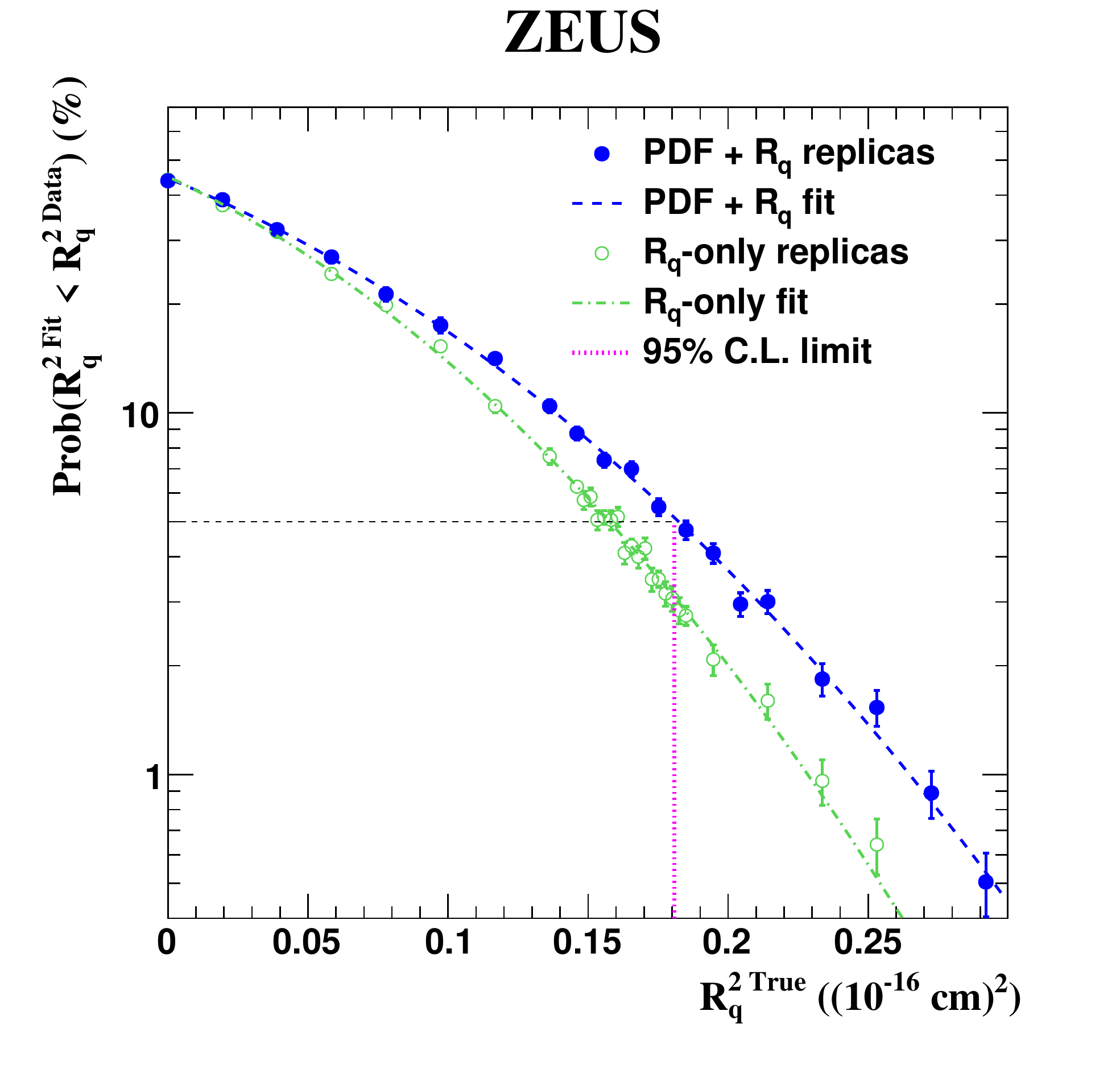}
\end{center}
\vspace*{-0.7cm}
\caption{
The probability of obtaining $R_{q}^{2 \; Fit}$ values smaller 
than that obtained for the actual data, $R_q^{2\; Data}$, calculated
from Monte Carlo replicas, as a function of the assumed value 
for the quark-radius squared, $R_{q}^{2\;True}$. 
Points with statistical error bars represent Monte Carlo replica sets
generated for different values of $R_{q}^{2\;True}$.
The solid circles correspond to the results obtained from the
simultaneous fit of $R_{q}^{2}$ and PDF parameters (PDF+$R_q$).
For comparison, the open circles represent the dependence obtained
when using the PDF parameters obtained from the QCD fit
neglecting BSM contribution ($R_q$-only). 
}
\label{fig-rq_prob_central}
\end{figure}

\begin{figure}[p]
\begin{center}
\includegraphics[width=0.54\textwidth]{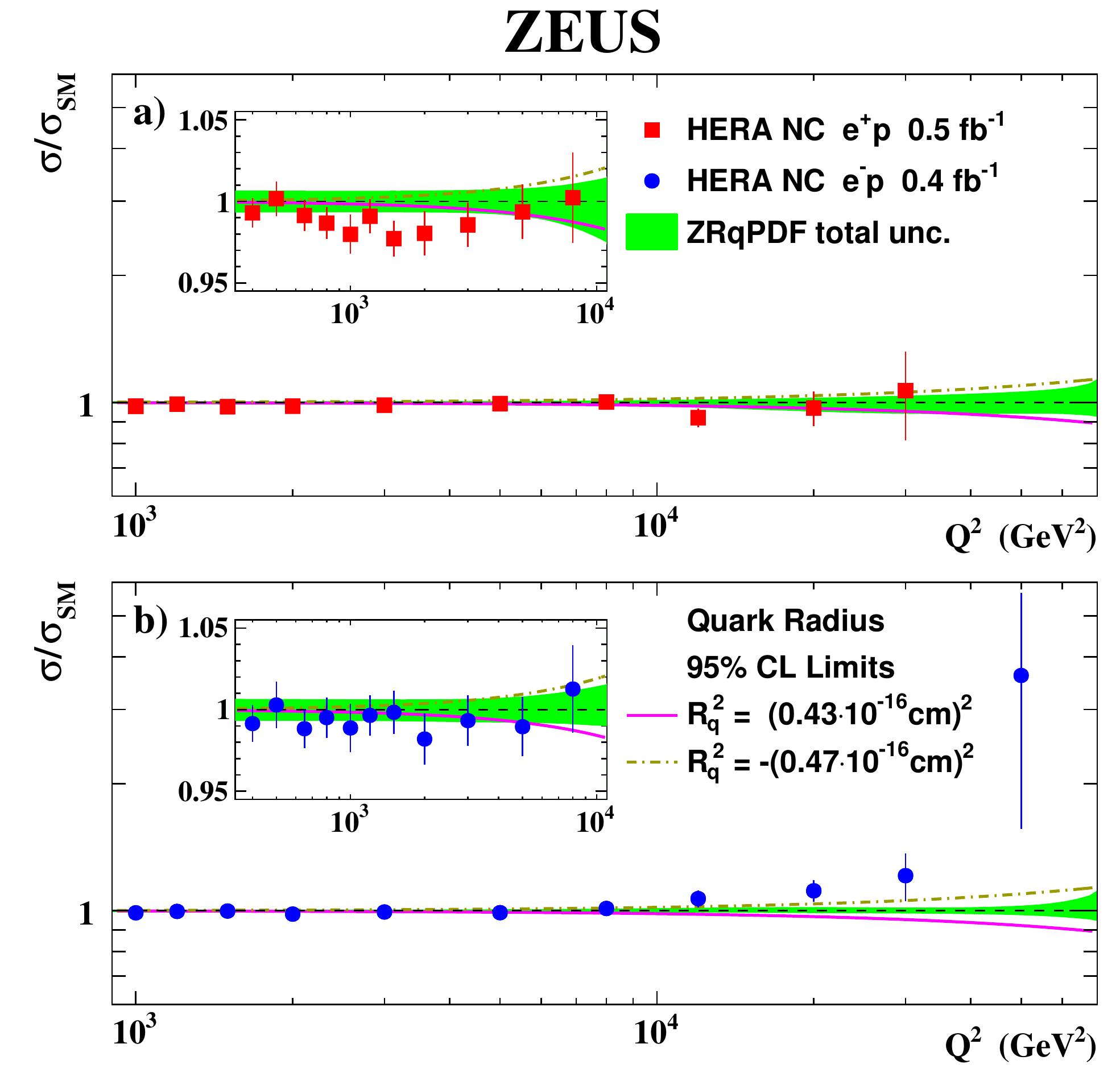}
\end{center}
\vspace*{-0.5cm}
\caption{
Combined HERA (a) $e^+p$ and (b) $e^-p$ NC DIS data 
compared to the $95\%$~C.L. exclusion limits on 
the effective mean-square radius of quarks. 
Cross sections are normalized to the Standard Model predictions
resulting from the QCD fit to combined HERA data, without BSM
contribution \protect\cite{zeus_rq_paper}. 
The bands represent the total uncertainty on the predictions. 
The insets show the comparison in the $Q^{2} < 10^{4}$~GeV$ \, {}^{2}$ 
region with a linear ordinate scale.
}
\label{fig-rq_limit}
\end{figure}

The results of the limit-setting procedure using the simultaneous fit to PDF
parameters and $R_{q}^{2}$, based on sets of Monte
Carlo replicas testing the possible cross-section modifications
due to a quark form factor, yield the $95\%$~C.L. limits on the effective 
quark radius of
\begin{eqnarray*}
  -(0.47\cdot 10^{-16} {\rm cm})^2 \; < \;
     R_q^{2} & < & (0.43\cdot 10^{-16} {\rm cm})^2 \; ,
\end{eqnarray*}  
see Fig.~\ref{fig-rq_limit}.
Taking into account the possible influence of quark radii on the PDF 
parameters is necessary as demonstrated in 
Fig.~\ref{fig-rq_prob_central}, because the limits that would be
obtained for fixed PDF parameters are too strong 
by about 10\%.

\begin{figure}[!ht]
\begin{center}
\includegraphics[width=0.54\textwidth]{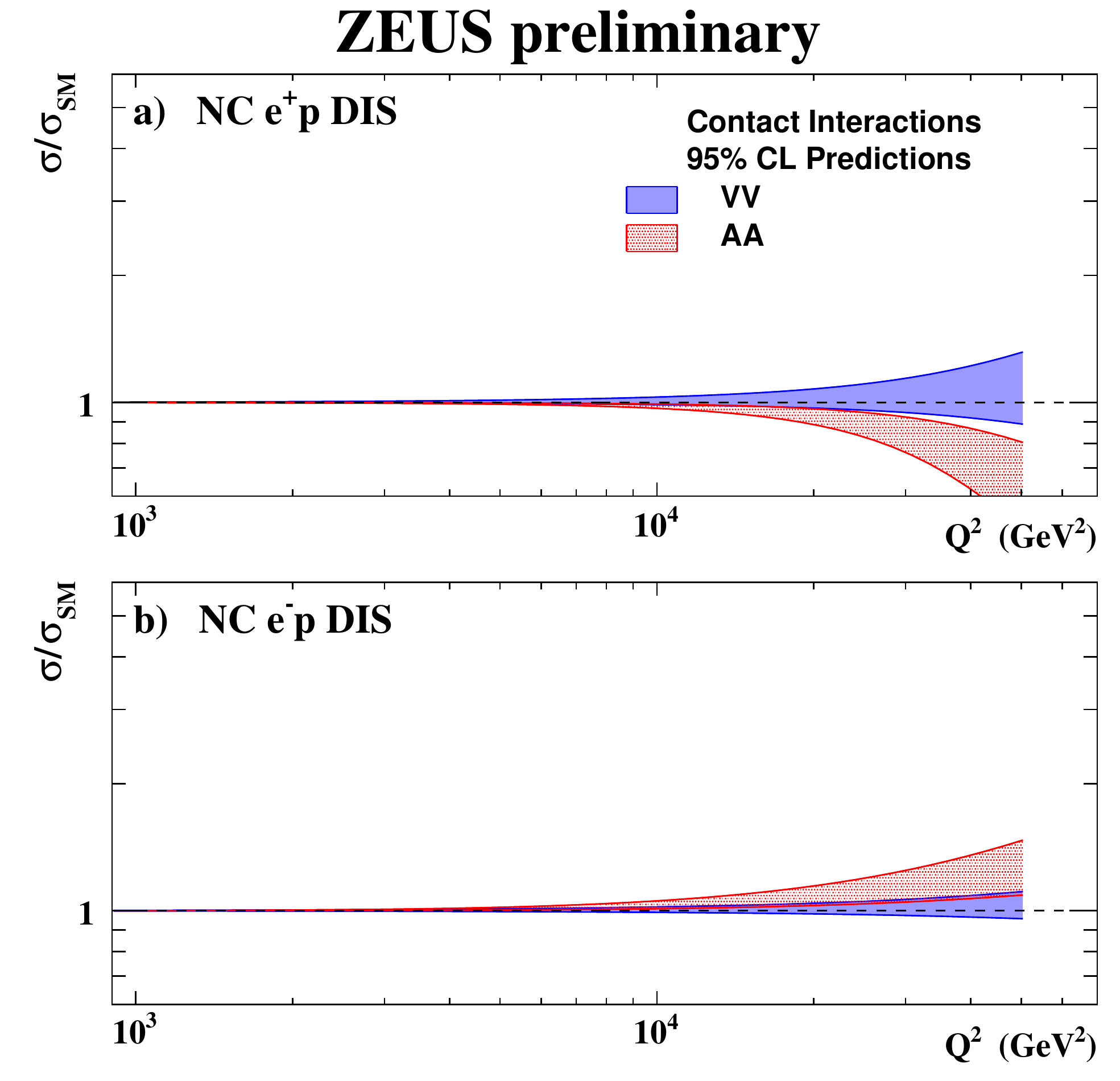}
\end{center}
\vspace*{-0.5cm}
\caption{
  Cross section deviations from the SM predictions  allowed at
  $95\%$~C.L. for (a) $e^+p$ and (b) $e^-p$ NC DIS, as resulting from 
the analysis of HERA combined data in the VV and AA
Contact Interaction scenarios.
}
\label{fig-vvaa_limit}
\end{figure}

Shown in Fig.~\ref{fig-vvaa_limit} are cross section deviations
from the SM predictions corresponding to the coupling range allowed at
$95\%$~C.L. for VV and AA contact-interaction models.
No significant deviation from the SM predictions is observed for
the VV scenario.
The probability of obtaining larger best-fit coupling 
for $\eta^{\rm True} = \eta^{\rm SM} = 0$, 
i.e. the probability that an experiment, assuming the validity of the SM,
would produce a value of $\eta^{\rm Fit}$ greater than that obtained
from the data, $\eta^{\rm Fit} > \eta^{\rm Data}$, is $p_{SM} =  25$\%. 
However, for the AA scenario the best $\chi^2$ value is for
$\eta^{\rm Data} = 0.32 \cdot 10^{-6}\;{\rm GeV}^{-2}$, corresponding to a
compositeness scale of about 6.2~TeV. The SM probability is
$p_{SM} = 0.7$\%, which corresponds to a deviation of about $2.5 \; \sigma$. 
Negative values of $\eta$ are excluded at $95\%$~C.L. for this scenario
and the allowed range for positive couplings corresponds to compositeness
scales $4.8 < \Lambda^{+} < 10.4$~TeV.

The $95\%$~C.L. limits on the compositeness scale $\Lambda$ 
for different CI models are summarised in Tab.~\ref{tab-ci}.
Also for the VA and X1 scenarios, an improved description of the HERA data
can be obtained.
The largest improvement is $\Delta \chi^2/$d.o.f. of -0.005 for the X1 model.
The probability that the SM experiment reproduces the data is
$p_{SM}=0.3\%$.
The fits suggest that the SM underestimates NC DIS cross sections
at the highest $Q^2$ for $e^{-}p$, while the predictions for $e^{+}p$ are
slightly too high.

%----------------------------------------------------------------------------
%      Figure: Limit setting procedure for Rq
%----------------------------------------------------------------------------

%-------------------------------------------------------------------------------
%      Figure: Rq results
%-------------------------------------------------------------------------------

%-------------------------------------------------------------------------------
%       Figure: CI limit plot
%-------------------------------------------------------------------------------

%------------------------------------------------------------------------------
%       CI model and result table
%------------------------------------------------------------------------------

\addtolength{\tabcolsep}{2mm}

\begin{table}[!ht]
  \begin{center}
   \begin{tabular}{|c@{~~~[}r@{,}r@{,}r@{,}r|cc|cc|r|}
  \multicolumn{10}{c}{{\bf ZEUS preliminary }} \\
  \multicolumn{10}{c}{{HERA $e^\pm p$ 1994-2007 data}} \\
\hline
  \multicolumn{5}{|c|}{}  &  
  \multicolumn{4}{c|}{{\bf $95\%$~C.L. limits (TeV)}} & \\
\cline{6-9}
  \multicolumn{5}{|c|}{Coupling structure}  &  
  \multicolumn{2}{c|}{Measured}     &
  \multicolumn{2}{c|}{Expected}     &
  $p_{SM}$ \\
Model  & 
 $\epsilon_{_{LL}}$ & $\epsilon_{_{LR}}$  &
                 $\epsilon_{_{RL}}$ &  $\epsilon_{_{RR}}$]~~ &
~~~$\Lambda^{-}$  &  $\Lambda^{+}$ &
~~~$\Lambda^{-}$  &  $\Lambda^{+}$ &
   (\%) \\
\hline
LL 
&  +1  &  0  &  0  &  0]~~
 & 22.0  & 4.5
 & 5.9 & 6.2 
 & 6.5 \\
RR 
 &  0  &  0  &  0  &  +1]~~      
 & 32.9  &  4.4
 & 5.7  & 6.1 
 & 5.6 \\
\hline
VV 
 &  +1  &  +1  &  +1  &  +1]~~    
 & 14.7  &  9.5
 & 11.0  & 11.4 
 &  24.8 \\
AA 
 &  +1  &  $-1$  &  $-1$  &  +1]~~ 
 &  - &  4.8 - 10.4
 & 7.9 & 7.8  
 & 0.7 \\
VA 
 &  +1  &  $-1$  &  +1  &  $-1$]~~ 
 & -  &  3.6 - 10.1
 & 4.1 & 4.1  
 & 2.1 \\
X1 
 &   +1  &  $-1$  &  0  &  0]~~ 
 & -  & 3.5 - ~6.6
 &  5.7 & 5.6 
 & 0.3 \\
X2 
 &   +1  &  0  &  +1  &  0]~~ 
 & 10.8 & 6.8
 & 7.8 & 8.2  
 & 23.1 \\
X4 
 &   0  &  +1  &  +1  &  0]~~ 
 & 7.6 & 9.2
 & 8.0 & 8.6
 & 60.3  \\
\hline
    \end{tabular}
  \end{center}
  \caption{
     Relations between couplings 
  $[\epsilon_{LL},\epsilon_{LR}, \epsilon_{RL}, \epsilon_{RR}]$ for the 
         compositeness models and the $95\%$~C.L. limits on
         the compositeness scale, $\Lambda$,
         obtained from the ZEUS analysis of the HERA inclusive data.
         Also shown are the expected limits,
         and the SM probability, $p_{SM}$.
         Each row of the table represents two scenarios corresponding to
         $\eta>0$ ($\Lambda^{+}$) and $\eta<0$ ($\Lambda^{-}$).
         Negative coupling values are excluded at $95\%$~C.L. for AA,
         VV and X1 models, and only the given range of compositeness
         scales for positive coupling sign is allowed. 
        }
  \label{tab-ci}
\end{table}

%---------------------------------------------------------------------

\end{document}